\documentclass[pra,twocolumn,superscriptaddress]{revtex4}
\usepackage{epsf}
\usepackage{amsmath}

\def\llangle{\langle\!\langle}
\def\rrangle{\rangle\!\rangle}

\begin{document}

\title{A Software Simulator for Noisy Quantum Circuits}
\author{Himanshu Chaudhary}
\affiliation{Physics Department, Indian Institute of Science,
             Bangalore 560012}
\author{B. Mahato}
\affiliation{Physics Department, Indian Institute of Science,
             Bangalore 560012}
\author{Lakshya Priyadarshi}
\affiliation{Computer Science and Engineering Department,
             Institute of Engineering and Technology, Lucknow 226021}
\author{Naman Roshan}
\affiliation{Physics Department, Indian Institute of Technology,
             Kanpur 208016}
\author{Utkarsh}
\affiliation{Center for Computational Natural Sciences and Bioinformatics,
             International Institute of Information Technology, Hyderabad 500032}
\author{Apoorva D. Patel}
\email{adpatel@iisc.ac.in}
\affiliation{Centre for High Energy Physics, Indian Institute of Science,
             Bangalore 560012}

\date{\today}

\begin{abstract}
We have developed a software library that simulates noisy quantum logic
circuits. We represent quantum states by their density matrices in the
Pauli basis, and incorporate possible errors in initialisation, logic
gates, memory and measurement using simple models. Our quantum simulator
is implemented as a new backend on IBM’s open-source Qiskit platform.
In this document, we provide its description, and illustrate it with
some simple examples.
\end{abstract}

\maketitle

\section{Motivation}

The field of quantum technologies has made rapid strides in recent years,
and is poised for significant breakthroughs in the coming years.
Practical applications are expected to appear first in sensing and metrology,
then in communications and simulations, then as feedback to foundations of
quantum theory, and ultimately in computation.
The essential features that contribute to these technologies are
superposition, entanglement, squeezing and tunneling of quantum states.
The theoretical foundation of the field is clear; laws of quantum mechanics
are precisely known, and elementary hardware components work as predicted
\cite{preskill,nielsen}.
The challenge is a large scale integration, say of 10 or more components.

Quantum systems are highly sensitive to disturbances from the environment;
even necessary controls and observations perturb them.
The available, and upcoming, quantum devices are noisy, and techniques
to bring down the environmental error rate are being intensively pursued.
At the same time, it is necessary to come up with error-resilient system
designs, as well as techniques that validate and verify the results.
This era of noisy intermediate scale quantum systems has been labeled NISQ
\cite{NISQ}.
Such systems are often special purpose platforms, with limited capabilities.
They roughly span devices with 10-100 qubits, 10-1000 logic operations,
limited interactions between qubits, and with no error correction since the
fault-tolerance threshold is orders of magnitudes away.

Worldwide, many universities and companies have developed software quantum
simulators for help in investigations of noisy quantum processors
\cite{qcsimulators}.
They are programs running on classical parallel computer platforms,
and can model and benchmark 10-50 qubit systems.
(It is not possible to classically simulate larger qubit systems due to
exponential growth of the Hilbert space size.)
For a user accessing a computer on the cloud and obtaining the output
of a program, it makes little difference whether there is a genuine quantum
processor at the other end or just a suitable software simulator.
For such an imitation, instead of using exact algebra, the simulator has
to be designed to mimic a noisy quantum processor.

A quantum computation may suffer from many sources of error: due to imprecise
initial state preparation, due to imperfect logic gate implementation, due to
disturbances to the data in memory, and due to error-prone measurements.
(It is safe to assume that the program instructions, which are classical,
are essentially error-free.)
So a realistic quantum simulator would have to include all of them with
appropriate probability distributions.
Additional features that can be included are restrictions on possible logic
operations and connectivity between the components, which would imitate what
may be the structure of a real quantum processor.
With such improvisations, the simulation results would look close to what
a noisy quantum processor would deliver, and one can test how well various
algorithms work with imperfect quantum components.
More importantly, one can vary the imperfections and the connectivity in the
software simulator to figure out what design for the noisy quantum processor
would produce the best results.

Quantum simulators serve an important educational purpose as well.
They are portable, and can be easily distributed over existing computational
facilities world wide.
They are therefore an excellent way to attract students to the field of
quantum technology, providing a platform to acquire the skills of
`programming' as well as `designing' quantum processors.
Programming a quantum processor is qualitatively different from the
classical experience of computer programming, and so the opportunity and
exposure provided by quantum simulators would be of vital importance in
developing future expertise in the field.
It is with such an aim that we have constructed a software library for
simulating noisy quantum logic circuits.
We provide the details in what follows.

\section{Implementation}

In the standard formulation of quantum mechanics, states are vectors in a
Hilbert space and evolve by unitary transformations,
$|\psi\rangle \rightarrow U|\psi\rangle$.
This evolution is deterministic, continuous and reversible.
It is appropriate for describing the pure states of a closed quantum system,
but is insufficient for describing the mixed states that result from
interactions of an open quantum system with its environment.

The most general description of a quantum system is in terms of its density
matrix $\rho$, which evolves according to a linear completely positive trace
preserving map known as the superoperator \cite{preskill,nielsen}.
For generic mixed states, $\rho$ is a Hermitian positive semi-definite matrix,
with $Tr(\rho)=1$ and $\rho^2\preceq\rho$, while for pure states,
$\rho=|\psi\rangle\langle\psi|$ and $\rho^2=\rho$.
The density matrix provides an ensemble description of the quantum system,
and so is inherently probabilistic, in contrast to the state vector description
that can describe individual experimental system evolution.
Nonetheless, it allows determination of the expectation value of any physical
observable, $\langle O\rangle = Tr(O\rho)$, which is defined as the average
result over many experimental realisations.

In its discrete form, a superoperator can be specified by its Kraus
operator-sum representation:
\begin{equation}
\label{krausevol}
\rho \rightarrow \sum_\mu M_\mu \rho M_\mu^\dagger ~,~~
\sum_\mu M_\mu^\dagger M_\mu = I ~.
\end{equation}
Unitary evolution, $\rho \rightarrow U\rho U^\dagger$, and orthogonal
projective measurement, $\rho \rightarrow \sum_k P_k \rho P_k$, are special
cases of this representation.
(Note that the projection operators satisfy $P_k=P_k^\dagger$, $\sum_k P_k=I$
and $P_k P_l = P_k \delta_{kl}$.)
Also, various environmental disturbances to the quantum system can be modeled
by suitable choices of $\{M_\mu\}$.

In going from a description based on $|\psi\rangle$ to the one based on
$\rho$, the degrees of freedom get squared.
This property is fully consistent with the Schmidt decomposition, which
implies that any correlation between the system and the environment can be
specified by modeling the environment using a set of degrees of freedom as
large as that for the system.
The squaring of the degrees of freedom is the price to be paid for the
flexibility to include all possible environmental effects on the quantum
system, and it slows down the performance of our quantum simulator.

We consider computational problems whose algorithms have already been
converted to discrete quantum logic circuits acting on a set of qubits.
We also assume that all logic gate instructions can be executed with a
fixed clock step.
In this framework, the computational complexity of the program is
specified by the number of qubits and the total number of clock steps.
Since the quantum state deteriorates with time due to environmental
disturbances, we reduce the total execution time by identifying
non-overlapping logic operations at every clock step and then
implementing them in parallel.

Our quantum simulator is an open-source software written in Python,
which is added as a new backend to IBM's Qiskit platform \cite{qiskit}.
That extends the existing Qiskit capability, while retaining the convenience
(e.g. portability, documentation, graphical interface) of the Qiskit format.
Being an open-source software platform, Qiskit is popular, and a variety of
quantum algorithms have been implemented using it \cite{coles}.
Its comparison with other quantum computing software packages, in terms of
features and performance, is also available \cite{larose}.
Our simulator, with a user guide, is available as a ``derivative work" of
Qiskit at \url{https://github.com/indian-institute-of-science-qc/}
{\tt qiskit-aakash}.

\subsection{The Quantum State}

We express the density matrix of an $n$-qubit quantum register in the
orthogonal Pauli basis that follows from the tensor product structure
of the Hilbert space:
\begin{equation}
\label{rhodef}
\rho = \sum_{i_1,i_2,\ldots,i_n} a_{i_1 i_2\ldots i_n}
(\sigma_{i_1}\otimes\sigma_{i_2}\otimes\ldots\otimes\sigma_{i_n}) ~.
\end{equation}
Here $i_1,i_2,\ldots,i_n\in\{0,1,2,3\}$, $\sigma_0\equiv I$, and
$a_{i_1 i_2\ldots i_n}$ are $4^n$ real coefficients encoded as an array.
All eigenvalues of the density matrix lie in the interval $[0,1]$,
and add up to $1$.
The normalisation $Tr(\rho)=1$ implies $a_{00\ldots0}=2^{-n}$.
The constraint $Tr(\rho^2)\le1$, which follows from $\rho^2\preceq\rho$,
implies $\sum_{i_1,i_2,\ldots,i_n} a_{i_1 i_2\ldots i_n}^2 \le 2^{-n}$.
We find this expression for the density matrix easier to work with,
because the orthogonality of the Pauli basis makes it easy to describe
various transformations of the density matrix as simple changes of the
coefficients, as described in the rest of the article.
(When the density matrix is expressed as a $2^n\times2^n$ complex Hermitian
matrix, the number of independent components remain the same. But the matrix
elements do not belong to an orthogonal set, and their transformations do not
have the same type of compact description.)

Quantum dynamics is linear in terms of $\rho$.
Moreover, we consider problems where all operations---logic gates,
errors and measurements---are local, i.e. act on only a few qubits.
Indeed, the Qiskit transpiler decomposes more complicated operations
into a sequence of one-qubit and two-qubit operations.
The tensor product structure of such operations is a non-trivial operator
on the addressed qubits and the identity operator on the rest of the qubits
(e.g. $I\otimes\ldots\otimes O_k\otimes\ldots\otimes I$ for an operation on
the $k^{\rm th}$ qubit).
Since the expression for the quantum register has the same tensor product
structure, Eq.~(\ref{rhodef}), the operation changes the Pauli matrix
factors corresponding to only the addressed qubits (e.g. $\sigma_{i_k}$)
and the coefficients change only for the associated subscripts
(e.g. $a_{\ldots i_k\ldots})$.
Such operations are efficiently implemented in the software using
linear algebra vector instructions, with explicit evaluation of
Eq.~(\ref{krausevol}).
We list density matrix transformations for some commonly used logic
gates in Appendix A.

We allow several options to initialise the density matrix:
The all-zero state is $2^{-n}(I+\sigma_3)^{\otimes n}$, the uniform
superposition state is $2^{-n}(I+\sigma_1)^{\otimes n}$, a state specified
by a binary string of 0's and 1's is mapped to a matching tensor product
of $\frac{1}{2}(I+\sigma_3)$ and $\frac{1}{2}(I-\sigma_3)$ factors, and a
custom density matrix can be read from a file listing the $4^n$ coefficients.
We also use the overlap, $Tr(\rho_1\rho_2)$, as a convenient measure of
closeness of two density matrices.

\subsection{The Logic Gates}

Generic unitary transformations acting on the density matrix belong to the
group $SU(2^n)$, since $\rho$ does not contain the overall unobservable
phase that $|\psi\rangle$ has.
It is well-known that any such transformation can be decomposed in to a
sequence of one-qubit and two-qubit logic gates.
The optimal choice for these elementary gates depends on what operations
are convenient to execute on the quantum hardware, but it is a small set
in any case.
We choose this select set to be the one-qubit rotations about the fixed
Cartesian axes (i.e. $x$, $y$ and $z$) and the two-qubit C-NOT gate,
which is suitable for most hardware implementations.

We assume that the program to be executed is available as a time-ordered
sequence of logic gate operations.
If that is not so, then a transpiler is needed to convert instructions
in a high-level language to a sequence of logic gate operations.
We also assume that the C-NOT gate can be applied between any two qubits
of a register.
If there are restrictions on qubit connectivity, then again it would be the
task of a transpiler to express the desired C-NOT gate as a sequence of
operations along an available qubit interaction route.

We follow the Qiskit convention in describing the logic gates.
A single qubit rotation by angle $\theta$ about the axis $\hat{n}$ is
$R_n(\theta) = e^{-i\hat{n}\cdot\vec{\sigma}\theta/2}$.
Then using Euler decomposition, any single qubit rotation is decomposed as:
\begin{eqnarray}
u3(\theta,\phi,\lambda) &=& e^{-i(\phi+\lambda)/2}
  \begin{pmatrix}
  \cos\frac{\theta}{2} & -e^{i\lambda}\sin\frac{\theta}{2} \cr
  e^{i\phi}\sin\frac{\theta}{2} & e^{i(\phi+\lambda)}\cos\frac{\theta}{2} \cr
  \end{pmatrix}
  \nonumber\\
&=& R_z(\phi)R_y(\theta)R_z(\lambda) ~.
\end{eqnarray}
We also use the Qiskit preprocessor and transpiler to simplify the quantum
logic circuit.
First, the preprocessor converts several of the commonly used quantum logic
gates to the select set of gates, e.g. the phase gates are expressed in terms
of $u1(\theta)=R_z(\theta)$, the Hadamard gate becomes
$H=i~u3(\frac{\pi}{2},0,\pi)$, and the Toffoli gate (C$^2$-NOT) becomes
a combination of C-NOT and phase gates.
Then the Qiskit transpiler optimises the quantum logic circuit, wherever
possible, by collapsing adjacent gates and by cancelling gates using
commutation rules \cite{optcircuit}.

Finally, the tensor product structure is used to embed the elementary logic
gates in operations for the full quantum register, as explained in Section 2.1.

\subsection{Projective Measurements}

Given the density matrix $\rho$, the expectation value of any Hermitian
operator $O$ is easily obtained by expressing it in the orthogonal Pauli basis,
\begin{equation}
\label{opdef}
O = \sum_{i_1,i_2,\ldots,i_n} b_{i_1 i_2\ldots i_n}
(\sigma_{i_1}\otimes\sigma_{i_2}\otimes\ldots\otimes\sigma_{i_n}) ~,
\end{equation}
and then evaluating the inner product,
\begin{equation}
\label{expval}
\langle O \rangle = Tr(O\rho) = 2^n \sum_{i_1,i_2,\ldots,i_n}
a_{i_1 i_2\ldots i_n} b_{i_1 i_2\ldots i_n} ~.
\end{equation}
Furthermore, the reduced density matrix with the degrees of freedom of the
$k^{\rm th}$ qubit summed over, $Tr_k(\rho)$, is specified by the $4^{n-1}$
coefficients $2 a_{i_1\ldots i_{k-1} 0 i_{k+1}\ldots i_n}$, since only the
terms containing $\sigma_0$ provide a non-zero partial trace.
This prescription can be repeated to reduce the density matrix over
as many qubits as desired.

Quantum measurement modifies the quantum state as well. In terms of the
complete set of orthogonal projection operators $\{P_k\}$, the outcome
$k$ is obtained with probability $Tr(P_k\rho)$, and the post-measurement
density matrix becomes $\sum_k P_k \rho P_k$.
The result is that the components orthogonal to the direction of
measurement vanish upon a projective measurement.
So when the $k^{\rm th}$ qubit is measured along direction $\hat{n}$,
the coefficients $a_{\ldots i_k \ldots}$ are set to zero for
$i_k\perp\hat{n}$, while those for $i_k\|\hat{n}$ and $i_k=0$
remain unchanged.

With these ingredients, and the orthogonality condition
$Tr(\sigma_i\sigma_j)=2\delta_{ij}$, we have implemented
several projective measurement options:\\
$\bullet$ {\it Single qubit measurement:}
When the $k^{\rm th}$ qubit is measured along direction $\hat{n}$,
the two possible results have probabilities
$\frac{1}{2}\langle(I\pm\hat{n}\cdot\vec{\sigma})_k\rangle$.
In terms of the coefficients $\{c_0,\vec{c}\} = a_{0 \ldots i_k \ldots 0}$,
the two probabilities are $\frac{1}{2}(1\pm 2^n\hat{n}\cdot\vec{c})$.
The density matrix is updated by projecting the coefficients with subscript
$i_k$.\\
$\bullet$ {\it Ensemble measurement:}
For simultaneous binary measurement of all the qubits in the computational
basis, the probabilities of the $2^n$ possible results are:\\
$2^{-n}\langle\prod_k(I\pm\sigma_3)_k\rangle
= \sum_{j_1,\ldots,j_n\in\{0,3\}} (\prod_k s_k) a_{j_1\ldots j_n}$,
with the sign $s_k=\delta_{j_k,0}\pm\delta_{j_k,3}$.
This measurement is typically performed at the end of the computation.\\
$\bullet$ {\it Expectation value of a Pauli operator string:}
This is just
$\langle \sigma_{i_1}\otimes\sigma_{i_2}\otimes\ldots\otimes\sigma_{i_n}
\rangle = 2^n a_{i_1 i_2\ldots i_n}$.
The density matrix is updated for each $k$, by projecting the coefficients
with subscripts $i_k\in\{1,2,3\}$ and leaving those with subscripts $i_k=0$
unaltered.\\
$\bullet$ {\it Bell-basis measurement of a pair of qubits:}
For this joint measurement of two qubits, the projection operators for
the four orthonormal state vectors are given by:
\begin{eqnarray}
\frac{|00\rangle+|11\rangle}{\sqrt{2}} \cdot
\frac{\langle00|+\langle11|}{\sqrt{2}} &=&
  \textstyle{\frac{1}{4}} (I\otimes I + \sigma_1\otimes\sigma_1
  - \sigma_2\otimes\sigma_2 + \sigma_3\otimes\sigma_3), \nonumber\\
\frac{|00\rangle-|11\rangle}{\sqrt{2}} \cdot
\frac{\langle00|-\langle11|}{\sqrt{2}} &=&
  \textstyle{\frac{1}{4}} (I\otimes I - \sigma_1\otimes\sigma_1
  + \sigma_2\otimes\sigma_2 + \sigma_3\otimes\sigma_3), \nonumber\\
\frac{|01\rangle+|10\rangle}{\sqrt{2}} \cdot
\frac{\langle01|+\langle10|}{\sqrt{2}} &=&
  \textstyle{\frac{1}{4}} (I\otimes I + \sigma_1\otimes\sigma_1
  + \sigma_2\otimes\sigma_2 - \sigma_3\otimes\sigma_3), \nonumber\\
\frac{|01\rangle-|10\rangle}{\sqrt{2}} \cdot
\frac{\langle01|-\langle10|}{\sqrt{2}} &=&
  \textstyle{\frac{1}{4}} (I\otimes I - \sigma_1\otimes\sigma_1
  - \sigma_2\otimes\sigma_2 - \sigma_3\otimes\sigma_3). \nonumber
\end{eqnarray}
So for the Bell-basis measurement of $k^{\rm th}$ and $l^{\rm th}$ qubits,
the probabilities of the four outcomes are determined by the coefficients
with subscripts $i_k=i_l$ and all other subscripts set to zero.
These four probabilities can be used to quantify entanglement between the
two qubits.
The post-measurement density matrix is obtained by setting the coefficients
with $i_k\ne i_l$ to zero, while not changing those with $i_k=i_l$.\\
$\bullet$ {\it Qubit reset:}
Although not a measurement, this instruction permits reuse of a qubit.
The transformation to reset a quantum state to $|0\rangle$ is:
$\rho \rightarrow P_0 \rho P_0 + \sigma_1P_1 \rho P_1\sigma_1$,
i.e. the projection along direction $|0\rangle$ is retained and the
projection along direction $|1\rangle$ is returned to direction $|0\rangle$.
When the $k^{\rm th}$ qubit is reset, the coefficients with $i_k\in\{1,2\}$
are made zero, and the coefficient with $i_k=3$ is made equal to the one
with $i_k=0$.\\
$\bullet$ {\it Complete tomography:} The $4^n$ coefficients of the density
matrix determine expectation values of all the physical operators that
can be measured in principle, although only a set of mutually commuting
operators can be measured in a single quantum experiment.
In our classical simulation, we can store the full density matrix at any
stage of a program, and use it later as the initial state of another
program.

\subsection{Computational Complexity}

Algorithmic instructions are decomposed by the Qiskit transpiler, which is a
type of compiler, into a sequence of elementary one- and two-qubit operations
that form a universal set.
In a quantum computer, these elementary operations would be directly executed
on quantum hardware, and the complexity of the algorithm is specified in terms
of the number of qubits and the number of elementary operations required.
In a simulator, the elementary operations are executed classically using
linear algebra, and so the complexity of a simulator can be specified in terms
of the linear algebra operations (permutations, additions and multiplications)
required to execute various elementary operations.
For a quantum state vector simulator and our density matrix simulator,
we can easily list the linear algebra complexity requirements for various
elementary operations as follows:\\
$\bullet$ {\it Memory:} Storage of an $n$-qubit register requires $2^n$
complex variables in the state vector representation and $4^n$ real
variables in the density matrix representation.\\
$\bullet$ {\it Logic gates:} With the tensor product structure, the local
one- and two-qubit operations can be expressed as block-diagonal matrices
with fixed block sizes. The linear algebra operations required for unitary
evolution are then $O(2^n)$ for the state vector representation and $O(4^n)$
for the density matrix representation.\\
$\bullet$ {\it Environmental noise:} For local environmental interactions,
the noise is parametrised in terms of a set of one-qubit Kraus operators,
which are also block-diagonal matrices with fixed block sizes.
Their effect cannot be evaluated in the state vector representation, but it
can be included in the density matrix representation with $O(4^n)$ linear
algebra operations.\\
$\bullet$ {\it Measurement:} The probability of a measurement outcome for
an observable is determined by the magnitude of the corresponding basis
component of the quantum state.
This is just a value look up in a simulator, and for an $n$-qubit register
it is an $O(n)$ effort in both the state vector and the density matrix
representations.

Overall, in going from a state vector simulator to a density matrix one,
the computational resource requirements increase from $O(2^n)$ to $O(4^n)$.
This change in scaling behaviour is due to increase in the number of
variables and would hold for any density matrix simulator.
It implies that with certain given computational resources, a density
matrix simulator can simulate half the number of qubits compared to a
state vector one.
That is the price paid for adding the capability to include generic local
environmental noise to the simulation.

\subsection{The Partitioned Logic Circuit}

An open quantum system continuously deteriorates in time.
To mitigate that, it is useful to reduce the total execution time of a
quantum program as much as possible.
Towards this end, we restructure the quantum circuit produced by the
Qiskit transpiler as follows.

The preprocessor decomposes the logic gates provided by the user to the
select set $\{u1,u3,$C-NOT$\}$.
This decomposition adds to the number of logic gates in the circuit,
increasing its depth.
As a countermeasure, we go through the sequence of operations on each qubit,
and merge consecutive single-qubit rotations that we find into a single one
(e.g. $u3*u3 \rightarrow u3$), using $SU(2)$ group composition rules.

Next we arrange the complete list of instructions in to a set of partitions,
such that all operations in a single partition can be executed as parallel
threads during a single clock step.
To accomplish this, the clock step has to be longer than the execution times
of individual operations (i.e. $u3$, C-NOT and various measurements).
We partition the circuit by organising the list of instructions as a stack
of sequential operations for every qubit, introducing barriers such that
each qubit can have at most one operation in a partition, and collecting
non-overlapping qubit operations into a single partition wherever possible.
In particular, this procedure puts logic gate operations and measurement
operations in separate partitions (note that a single qubit measurement
may affect the whole quantum register in case of entangled quantum states).
Thus a partition may have either multiple quantum logic gates operating on
different qubits, or multiple single qubit measurements on distinct qubits.
We let expectation value calculations, ensemble measurements and Bell-basis
measurements form partitions on their own.

We provide details of the merging and the partitioning logic circuit
operations in Appendices A and B, with simple illustrations.

\section{The Noisy Evolution}

The manipulations of circuit operations described in the previous section
are carried out at the classical level; even when a quantum hardware is
available, they would be implemented by a classical compiler.
So we safely assume that they are error-free.
It is the execution of the partitioned circuit on a quantum backend that
is influenced by the environment.
Assuming that the environment disturbs each qubit independently, we now
present simple models that include the environmental noise in the simulator
at various stages of the program execution.

\subsection{Initialisation Error}

The initial state of the program is often an equilibrium state.
So we allow a fully-factorised thermal state as one of the initial state
options:
\begin{equation}
\rho_{\rm th} = \begin{pmatrix}
                p & 0 \cr 0 & 1-p \cr
                \end{pmatrix}^{\otimes n} ,~~
\frac{p}{1-p} = \exp\left(\frac{E_1-E_0}{kT}\right) .
\end{equation}
Here the parameter $p$ is provided by the user.

\subsection{Logic Gate Execution Error}

The single qubit rotations in our select set have fixed rotation axes,
and we assume that errors arise from inaccuracies in their rotation angles.
Let $\alpha$ denote the inaccuracy in the angle, with the mean
$\llangle\alpha\rrangle=\overline{\alpha}$ and the fluctuations symmetric
about $\overline{\alpha}$.
Then the replacement $\theta\rightarrow\theta+\alpha$ in $R_n(\theta)$
modifies the density matrix transformation according to the substitutions:
\begin{equation}
\cos\theta \rightarrow r\cos(\theta+\overline{\alpha}) ~,~~
\sin\theta \rightarrow r\sin(\theta+\overline{\alpha}) ~,
\end{equation}
where $\overline{\alpha}$ and
$r = \llangle\cos(\alpha-\overline{\alpha})\rrangle$
are the parameters provided by the user.
They may depend on the rotation axis (i.e. $x$, $y$ or $z$).

To model the error in the C-NOT gate, we assume that C-NOT is implemented
as a transition selective pulse that exchanges amplitudes of the two
target qubit levels when the control qubit state is $|1\rangle$.
Then the error is in the duration of the transition selective pulse,
and alters only the second half of the unitary operator,
$U_{cx} = |0\rangle\langle0| \otimes I + |1\rangle\langle1| \otimes \sigma_1$.
It can be included in the same manner as the error in single qubit
rotation angle (i.e. as a disturbance to the rotation operator $\sigma_1$).
The corresponding two parameters, analogous to $\overline{\alpha}$ and
$r$, are provided by the user.

\subsection{Measurement Error}

Projective measurements of quantum systems are not perfect in practice.
We model a single qubit measurement error as depolarisation,
which is equivalent to a bit-flip error in a binary measurement.
Then when the $k^{\rm th}$ qubit is measured along direction $\hat{n}$,
the coefficients $a_{i_1 \ldots i_k \ldots i_n}$ in the post-measurement
state are set to zero for $i_k\perp\hat{n}$, reduced by a multiplicative
factor $d_1$ for $i_k\|\hat{n}$, and left unaffected for $i_k=0$.
Also, the probabilities of the two outcomes become
$\frac{1}{2}(1\pm 2^n d_1\hat{n}\cdot\vec{c})$,
in the notation of Section II.C.
Here the parameter $d_1$ is provided by the user.
In case of a measurement of a multi-qubit Pauli operator string, the above
procedure is applied to every qubit whose measurement operator has $i_k\ne0$.

In the case of a Bell-basis measurement, the post-measurement coefficients
with $i_k\ne i_l$ are set to zero, those with $i_k=i_l\in\{1,2,3\}$ are
reduced by a multiplicative factor $d_2$, and those with $i_k=i_l=0$ are
left the same.
Also, the probabilities of the four outcomes are obtained by reducing the
$i_k=i_l\in\{1,2,3\}$ contributions by the factor $d_2$ that is provided
by the user. 

\subsection{Memory Errors}

An open quantum system undergoes decoherence and decay, irrespective of
whether it is being manipulated by some instruction or not.
These effects cause maximum damage to a quantum signal, because they act
on all the qubits all the time, while operational errors are confined to
particular qubits at specific times.
We assume that these memory errors are small during a clock step, and
implement them by modifying the density matrix at the end of every clock
step, in the spirit of the Trotter expansion.
Such an implementation is actually the reason behind our partitioning of
the quantum circuit.

Taking the $\sigma_3$ basis as the computational basis, the decoherence
effect is to suppress the off-diagonal coefficients with $i_k\in\{1,2\}$
for every qubit by a multiplicative factor $f$.
It can be represented by the Kraus operators:
\begin{equation}
M_0 = \sqrt{\frac{1+f}{2}}~I ~,~~ M_1 = \sqrt{\frac{1-f}{2}}~\sigma_3 ~.
\end{equation}
In terms of the clock step $\Delta t$ and the decoherence time $T_2$,
the parameter $f=\exp(-\Delta t/T_2)$, and it is provided by the user.

We also consider the decay of the quantum state towards the thermal state,
$\rho_{\rm th}$, defined in Section III.A.
This evolution is represented by the Kraus operators:
\begin{eqnarray}
M_0 &=& \sqrt{p}\begin{pmatrix}
                1 & 0 \cr 0 & \sqrt{g} \cr\end{pmatrix} ,~
M_1  =  \sqrt{p}\begin{pmatrix}
                0 & \sqrt{1-g} \cr 0 & 0 \cr\end{pmatrix} ,\\
M_2 &=& \sqrt{1-p}\begin{pmatrix}
                  \sqrt{g} & 0 \cr 0 & 1 \cr\end{pmatrix} ,~
M_3  =  \sqrt{1-p}\begin{pmatrix}
                  0 & 0 \cr \sqrt{1-g} & 0 \cr\end{pmatrix} .\nonumber
\end{eqnarray}
Its effect on every qubit is to suppress the off-diagonal coefficients
with $i_k\in\{1,2\}$ by $\sqrt{g}$, and change the diagonal coefficients
according to:
\begin{equation}
a_{\ldots3\ldots} \rightarrow
g~a_{\ldots3\ldots} + (2p-1)(1-g) a_{\ldots0\ldots} ~.
\end{equation}
In terms of the clock step $\Delta t$ and the relaxation time $T_1$,
the parameter $g=\exp(-\Delta t/T_1)$, and it is provided by the user.
(Note that our Kraus representation automatically ensures the physical
constraint $T_2\le2T_1$).

The decoherence and decay superoperators commute with each other,
and we execute the combined operation at the end of every partition.

\section{Tests and Examples}

Several publicly available simulators work in the state vector formalism,
and obtain probabilistic results with multiple runs of the program (labeled
shots) \cite{larose}.
This formalism cannot handle mixed quantum states, and so cannot include
generic environmental noise.
Some simulators (e.g. Qiskit and Cirq) offer an option to work in the
density matrix formalism, but they represent the density matrix as a complex
number array.
Compared to them, our implementation is simpler due to the choice of the
Pauli basis to represent the density matrix.
The coefficients $a_{i_1i_2\ldots i_n}$ and $b_{i_1i_2\ldots i_n}$,
appearing in Eq.~(\ref{rhodef}) and Eq.~(\ref{opdef}) respectively,
are all real with this choice.
That completely eliminates complex number algebra from our code, and makes
the execution of linear algebra vector instructions more efficient.
Apart from this simulation advantage, it has also been observed that the
Pauli basis decomposition of the density matrix is highly efficient for
real quantum hardware measurements \cite{optbasis}. That follows from the
fact that all the Pauli basis elements,
$\sigma_{i_1}\otimes\sigma_{i_2}\otimes\ldots\otimes\sigma_{i_n}$,
mutually either commute or anticommute.

Our simulator code is written in Python, and it uses the software package
{\tt numpy} for execution of the linear algebra vector instructions.
Software packages such as {\tt numpy} are typically written in low level
langauges, and are tuned to the hardware used for execution.
Efficiency of a program written in a high level language (e.g. Python)
depends on how good these packages are, and the packages are often improved
as new types of hardware appear.
For example, {\tt numpy} can be replaced by {\tt JAX} for running the
program on multi-processor GPU's, and that requires just changing one
assignment statement at the beginning of our code.

We have tested our density matrix simulator against Qiskit's state vector
version, using circuits of randomly generated quantum logic operations.
Both give identical results, when all the errors are absent.
When errors are present, their accumulation during the whole circuit
program depends on the way the logic operations are organised.
There are differences in the organisation of logic operations between
our density matrix simulator and Qiskit's density matrix simulator.
So we have compared the noisy transformations generated for specific
types of errors in individual logic operations, between ours and
Qiskit's density matrix simulators, and those results have agreed.

Since our density matrix simulator works with $4^n$ coefficients, it is
slower than the state vector simulator that works with $2^n$ coefficients.
On the other hand, it produces the complete output probability distribution
in one run, while the state vector simulators require multiple runs of the
program for the same purpose.
We can simulate circuits with $10$ qubits and $100$ operations in a few
minutes on a laptop; it would be practical to handle larger quantum systems,
say up to $15$ qubits and $1000$ operations, on more powerful dedicated
computers.

The main achievement of our simulator is the ability to simulate noisy
quantum systems, using simple error models.
In such simulations, the final results are probability distributions over
the possible outcomes of the algorithm, and their stability against
variations of the error parameters can be explicitly checked.
The distributions can be easily visualised using various types of plots,
and we expect exponential deterioration of the quantum signal with
increasing error rates.

\begin{figure}
{\epsfxsize=8.6truecm \epsfbox{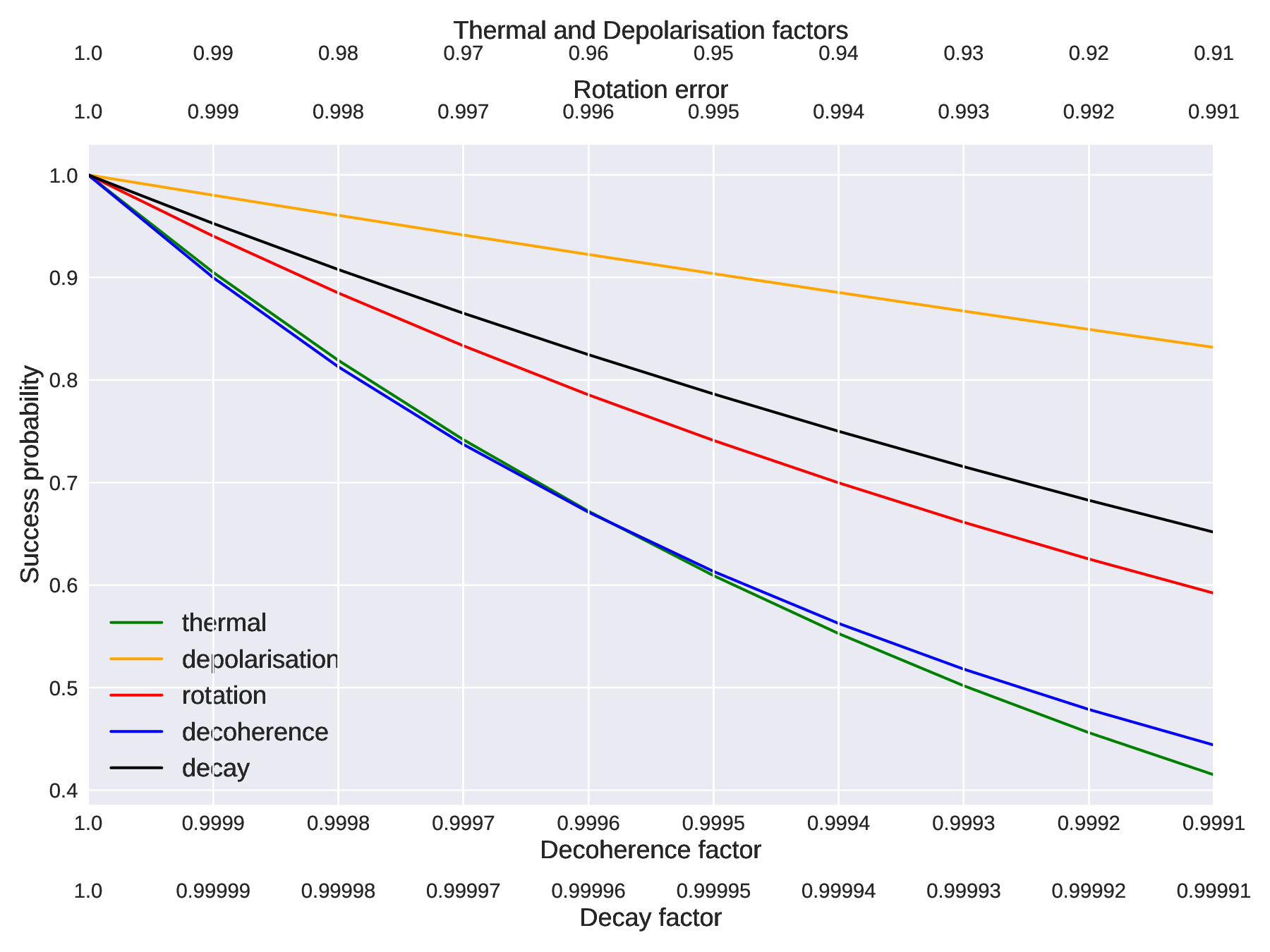}}
\caption{Success probability of the binary addition program,
110+11=1001, as a function of different types of errors.
For easy comparison, multiple parameters are plotted along the X-axis
with different scales: thermal factor $p$ (green), depolarisation factor
$d_1$ (orange), rotation error parameter $r$ (red), decoherence factor
$f$ (blue) and decay factor $g$ (black).
In all cases, the success probability is observed to decrease
exponentially.}
\label{err_comp}
\end{figure}

As a straightforward example, we simulated the binary addition algorithm.
That requires three quantum sub-registers, two for the two numbers and one
for the carry bits.
We varied the error parameters one at a time, and observed the probability
distributions of the final sum.
(To interpret the results correctly, we needed to invert Qiskit's convention
of the least significant bit first and the most significant bit last, to the
standard numerical convention of the most significant bit first and the least
significant bit last.)
Our results for the probability of the correct answer, for the addition
$110+11=1001$, are shown in Fig.~\ref{err_comp}.
Although the success probability decreases exponentially in all the cases,
we see a wide variation in sensitivity of the calculation to the different
types of errors.
Initialisation error parametrised in terms of the thermal factor ($p$), and
measurement error parametrised in terms of the depolarisation factor ($d_1$),
appear only at the two ends of the simulation, and cause the smallest changes
in the results.
The errors caused by rotation angle fluctuations ($r$) in the logic gates
produce intermediate size deviations in the results.
The errors represented by the decoherence factor ($f$) and the decay factor
($g$) act throughout the simulation, and cause the largest disturbances in
the results, with decay dominating over decoherence.
This pattern, together with the actual error parameter values, gives us
an estimate of how accurately we need to control various errors in quantum
hardware in order to get meaningful results.
We also observed that decreasing $p$, $d_1$ and $r$ more or less kept the
probability distribution centred around the correct answer, but decreasing
$f$ tended to make the distribution flat and decreasing $g$ drove the
distribution towards the all-zero state.

\begin{figure}
{\epsfxsize=8.5truecm \epsfbox{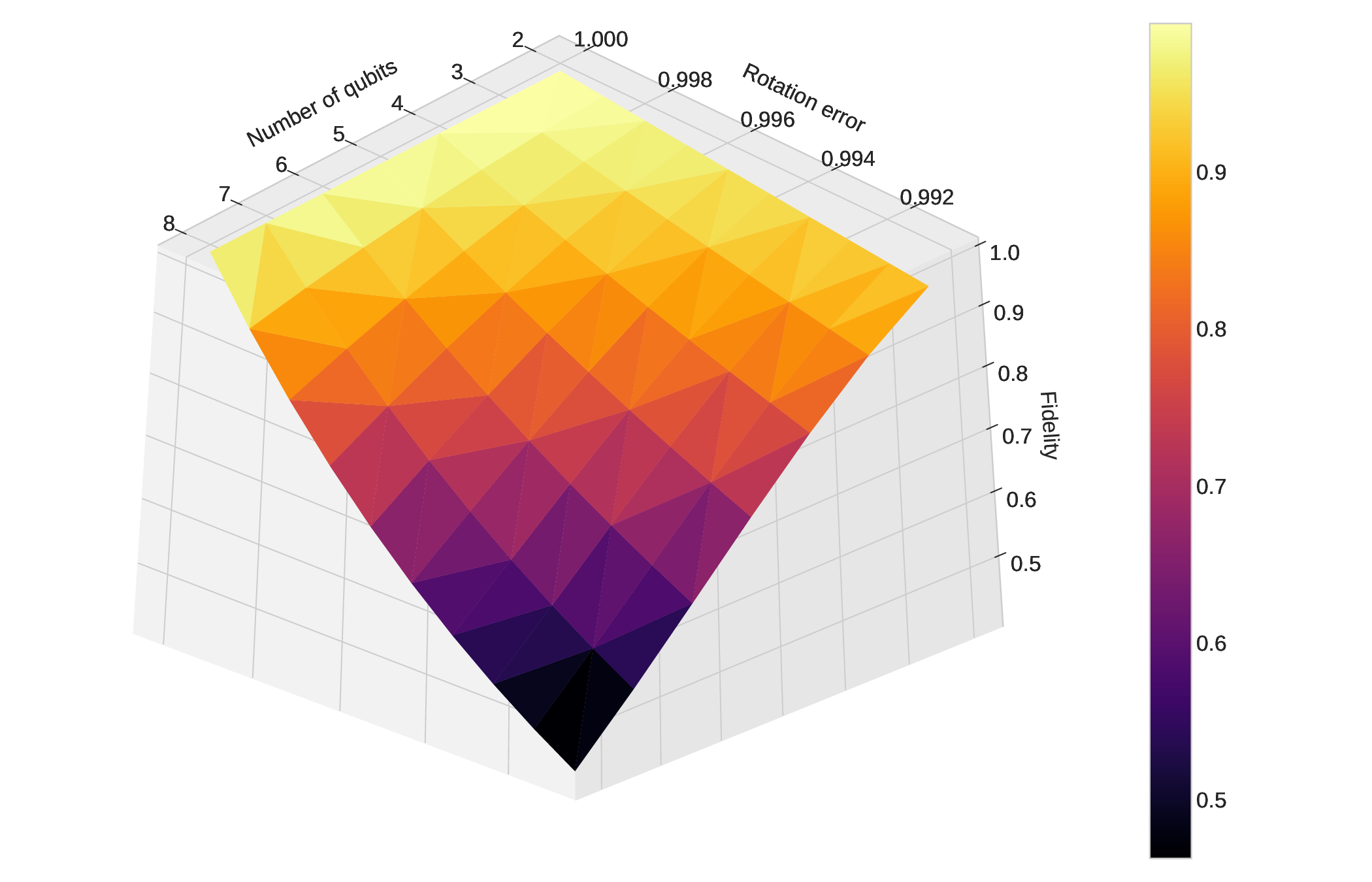}}
\null\vspace{-6mm}
{\epsfxsize=8.5truecm \epsfbox{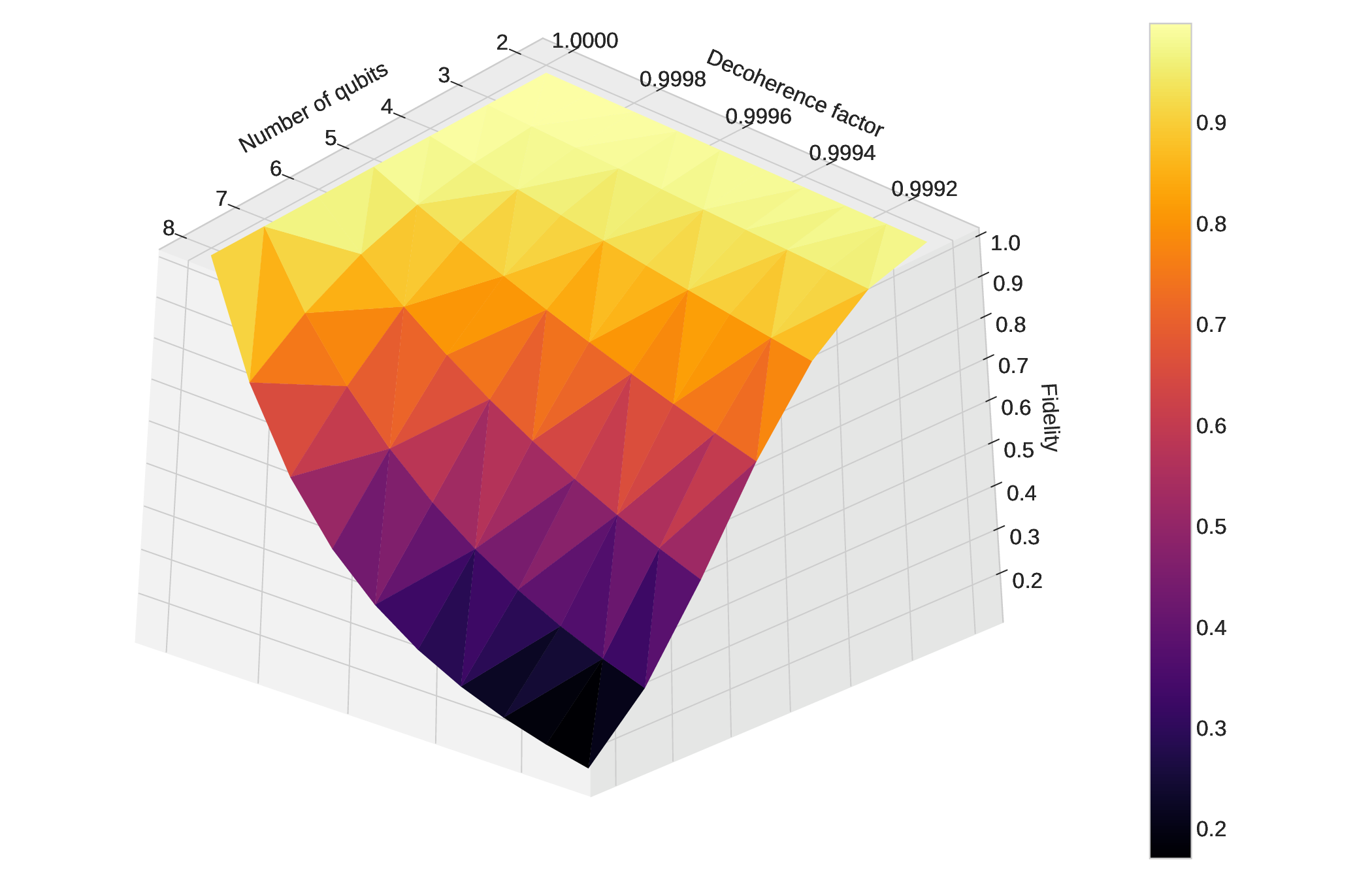}}
\null\vspace{-6mm}
{\epsfxsize=8.5truecm \epsfbox{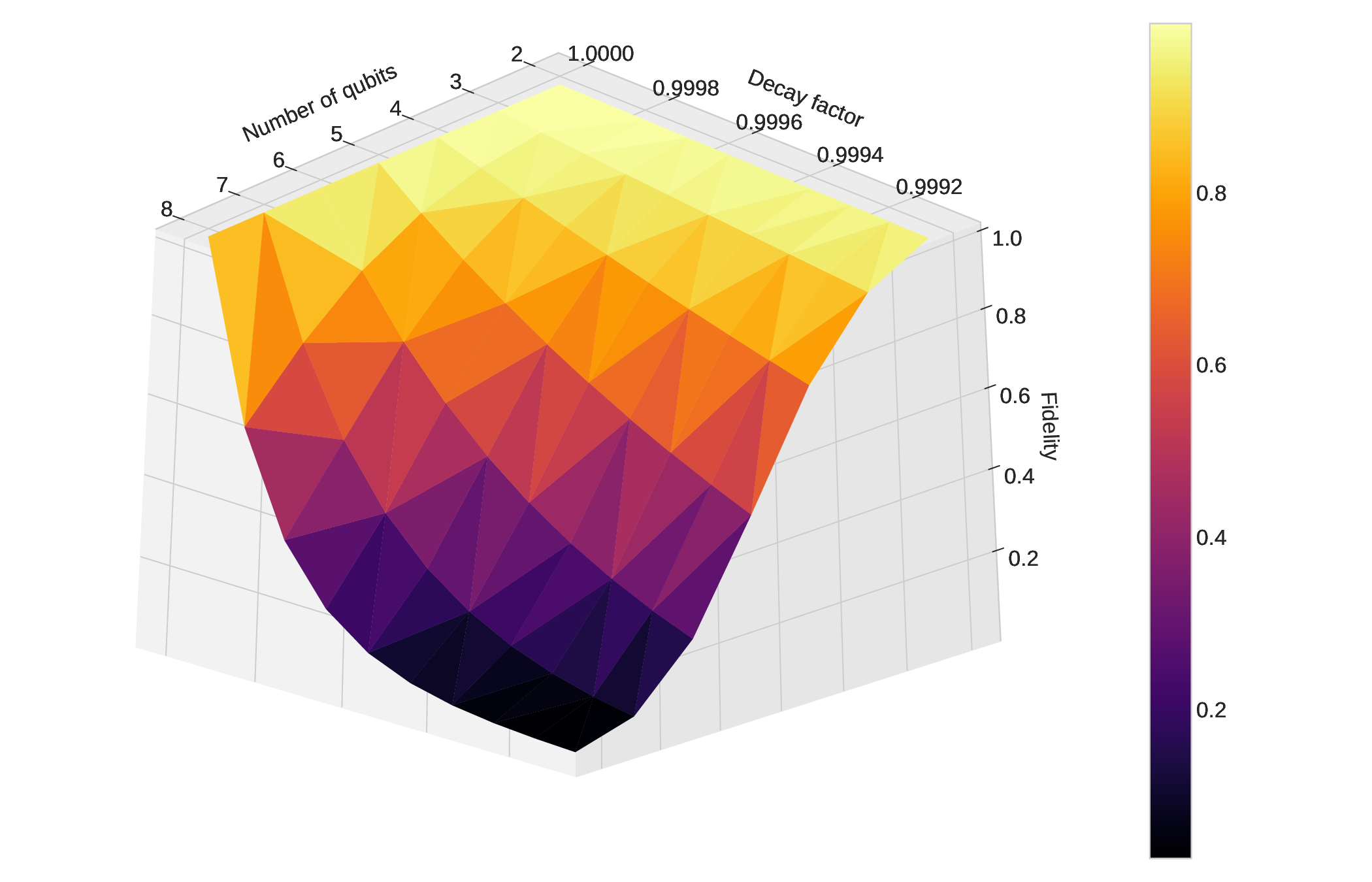}}
\caption{Fidelity of the Quantum Fourier Transform program, for different
number of qubits $n$ and as a function of different error parameters:
(Top) The rotation error parameter $r$, with the same value
for $R_x$, $R_y$, $R_z$ and C-NOT operations;
(Middle) The decoherence parameter $f$;
(Bottom) The decay parameter $g$.
The fidelity decreases first quadratically and then exponentially,
both as a function of $n$ and the error parameters.}
\label{qft_err}
\end{figure}

As a second example, we simulated the Quantum Fourier Transform algorithm
for the number of qubits ranging from $2$ to $8$, again varying the error
parameters one at a time.
Our results for the final state fidelity, with reference to the exact result,
are displayed in Fig.~\ref{qft_err} (the fidelity, $Tr(\rho_1\rho_2)$, is
evaluated the same way as in Eq.~(\ref{expval})).
We find that the fidelity deviates from $1$ quadratically for very small
errors, but subsequently drops exponentially, both as a function of the
number of qubits and the error parameters.
The hierarchy of sensitivity of the results to different errors is the same
as in case of the addition algorithm.

These observations are consistent with general theoretical expectations.
The errors that accumulate for the longest duration obviously alter the
results by the maximum amount.
Also, when the errors arise from an underlying unitary evolution, the change
in the results is a quadratic function of the error parameters when they are
very small (fidelity behaves as the cosine function).
On the other hand, the evolution is no longer unitary for larger values of
the error parameters, and the deterioration of the results is exponential.
In addition to confirming these expected scaling patterns, the simulations
provide information about the magnitudes of the resultant deviations, and
that is where their advantage and usefulness lie.

\section*{Acknowledgment}

This work was supported in part by the Department of Science and Technology,
Government of India, under the SERB grant EMR/2016/006312.

\section*{Appendix A: Some Logic Gate Transformations\\ for the Density Matrix}

The single qubit density matrix is $\rho=a_0I+\vec{a}\cdot\vec{\sigma}$,
with $a_0=\frac{1}{2}$.
It is straightforward to apply commonly used one-qubit logic gates to it:
\begin{eqnarray}
\sigma_1\rho\sigma_1 &=& a_0I + a_1\sigma_1 - a_2\sigma_2 - a_3\sigma_3 ~,
\nonumber\\
\sigma_2\rho\sigma_2 &=& a_0I - a_1\sigma_1 + a_2\sigma_2 - a_3\sigma_3 ~,
\nonumber\\
\sigma_3\rho\sigma_3 &=& a_0I - a_1\sigma_1 - a_2\sigma_2 + a_3\sigma_3 ~,
\nonumber\\
H\rho H &=& a_0I + a_3\sigma_1 + a_2\sigma_2 + a_1\sigma_3 ~,\\
S\rho S^\dagger &=& a_0I - a_2\sigma_1 + a_1\sigma_2 + a_3\sigma_3 ~,
\nonumber\\
S^\dagger\rho S &=& a_0I + a_2\sigma_1 - a_1\sigma_2 + a_3\sigma_3 ~,
\nonumber\\
T\rho T^\dagger &=& a_0I + \frac{a_1-a_2}{\sqrt{2}}\sigma_1
      + \frac{a_1+a_2}{\sqrt{2}}\sigma_2 + a_3\sigma_3 ~, \nonumber\\
T^\dagger\rho T &=& a_0I + \frac{a_1+a_2}{\sqrt{2}}\sigma_1
      - \frac{a_1-a_2}{\sqrt{2}}\sigma_2 + a_3\sigma_3 ~. \nonumber
\end{eqnarray}
Rotation errors in these one-qubit transformations are incorporated by
changing them from $R_n(\theta)\rho R_n^\dagger(\theta)$ to
$R_n(\theta+\alpha)\rho R_n^\dagger(\theta+\alpha)$.

The two-qubit C-NOT transformation, with the first qubit as the control,
is $U_{cx}\rho U_{cx}^\dagger$, where
$U_{cx} = |0\rangle\langle0|\otimes I + |1\rangle\langle1|\otimes\sigma_1$.
Transition selective pulse error in the C-NOT gate is included by changing
$\sigma_1$ to $R_x(\alpha)\sigma_1$ in $U_{cx}$.

Consecutive rotations of a single qubit can be merged into a single one
using $SU(2)$ group composition rules.
We rewrite Qiskit's $u2(\phi,\lambda)$ logic gate as
$u3(\frac{\pi}{2},\phi,\lambda)$, which reduces merging possibilities
to only four cases: $u1*u1$, $u1*u3$, $u3*u1$ and $u3*u3$.
The first three are easily taken care of by adding the $R_z$ rotation
angles, e.g. $u1(\theta_1) \times u1(\theta_2) = u1(\theta_1+\theta_2)$.
To take care of the last one, we express:
\begin{eqnarray}
& & u3(\theta_1,\phi_1,\lambda_1) * u3(\theta_2,\phi_2,\lambda_2) \\
&=& R_z(\phi_2)R_y(\theta_2)R_z(\lambda_2)
    R_z(\phi_1)R_y(\theta_1)R_z(\lambda_1) \nonumber\\
&=& R_z(\phi_2)R_y(\theta_2)R_z(\lambda_2+\phi_1)R_y(\theta_1)R_z(\lambda_1)
\nonumber\\
&=& R_z(\phi_2)R_z(\alpha)R_y(\beta)R_z(\gamma)R_z(\lambda_1) \nonumber\\
&=& R_z(\phi_2+\alpha)R_y(\beta)R_z(\gamma+\lambda_1) \nonumber\\
&=& u3(\beta,\phi_2+\alpha,\gamma+\lambda_1) ~. \nonumber
\end{eqnarray}
Here the YZY Euler decomposition on the third line is converted to the
ZYZ Euler decomposition on the third line.
This conversion is conveniently performed by explicitly matching the
product matrices and using the {\tt arctan2} math-library function to
extract the angles.

\begin{figure}
{\epsfxsize=8.6truecm \epsfbox{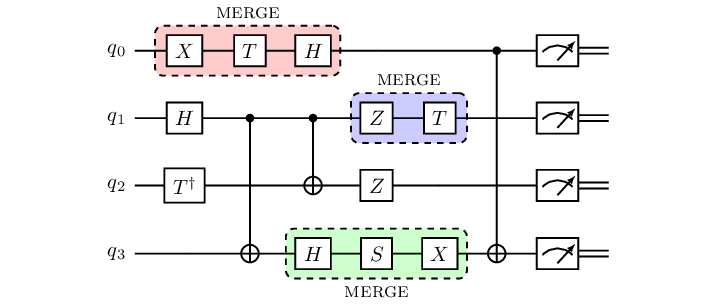}}
\null\vspace{2mm}
{\epsfxsize=8.6truecm \epsfbox{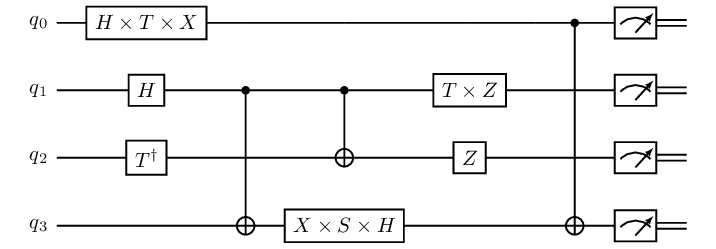}}
\caption{(Top) A quantum logic circuit with commonly used gates.
(Bottom) The same logic circuit after merging several one-qubit gates.}
\label{circ_merg}
\end{figure}

An illustration of how this merging can simplify a logic circuit is
presented in Fig.~\ref{circ_merg}. Note that the reversal of the operator
order is due to the convention of the leftmost gate acting first in a
circuit and the rightmost factor acting first in a matrix operation.

\section*{Appendix B: Logic Circuit Rearrangement}

To minimise decay and decoherence errors, we need to reduce the logic
circuit depth as much as possible.
For this purpose, we look for maximum parallelisation of the program
provided as a time-ordered instruction set.
We rearrange instructions in to a set of partitions, preserving their
temporal order, such that all instructions in a given partition commute
with each other and can be executed simultaneously while the partitions
are executed in succession.
Then each partition is assigned a clock step, and overall decay and
decoherence errors depend on the total number of partitions.

\begin{figure}
{\epsfxsize=8.6truecm \epsfbox{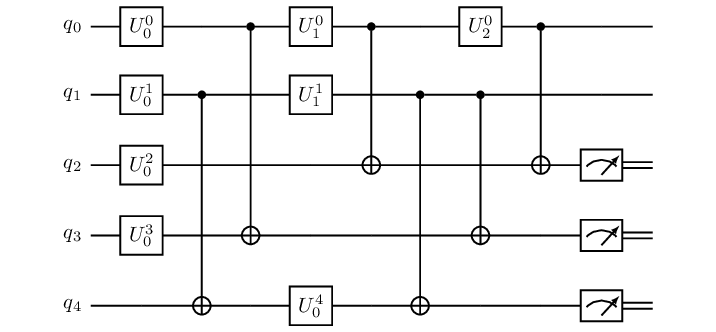}}
\null\vspace{2mm}
{\epsfxsize=8.6truecm \epsfbox{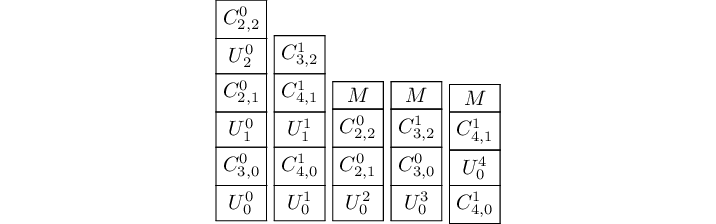}}
\null\vspace{2mm}
{\epsfxsize=8.6truecm \epsfbox{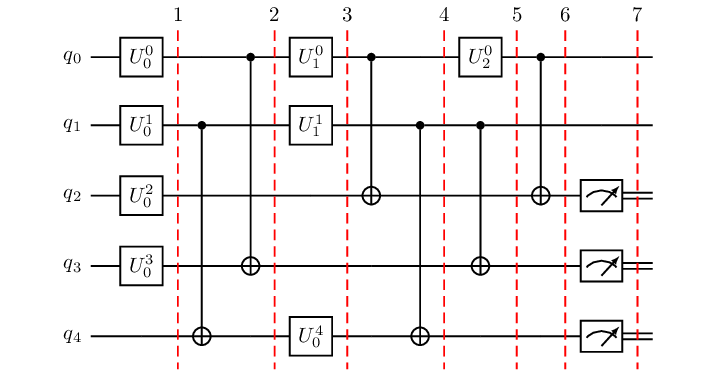}}
\caption{(Top) A quantum logic circuit specified by sequential instructions.
(Middle) The qubit operation stack generated from the logic circuit.
(Bottom) The partitioned logic circuit constructed from the qubit operation
stack.}
\label{circ_part}
\end{figure}

We note that all our logic gates including their errors involve only one
or two qubits, while any projective measurement operation may affect the
density matrix globally.
So the partitions fall in to two categories; they have either only unitary
logic gates ($u1$, $u3$ or C-NOT) or only projective measurement operations.
These two categories can have different clock step duration if required.
To begin with, we therefore go through the whole instruction set and insert
barriers between sets of consecutive logic gates and sets of consecutive
measurement operations.
We also separate out expectation value calculations, ensemble measurements
and Bell-basis measurements from the rest of the instructions by inserting
barriers.
The instructions between successive barriers are then inspected to check
if their further partitioning is necessary.

\begin{figure}[!h]
{\epsfxsize=8.6truecm \epsfbox{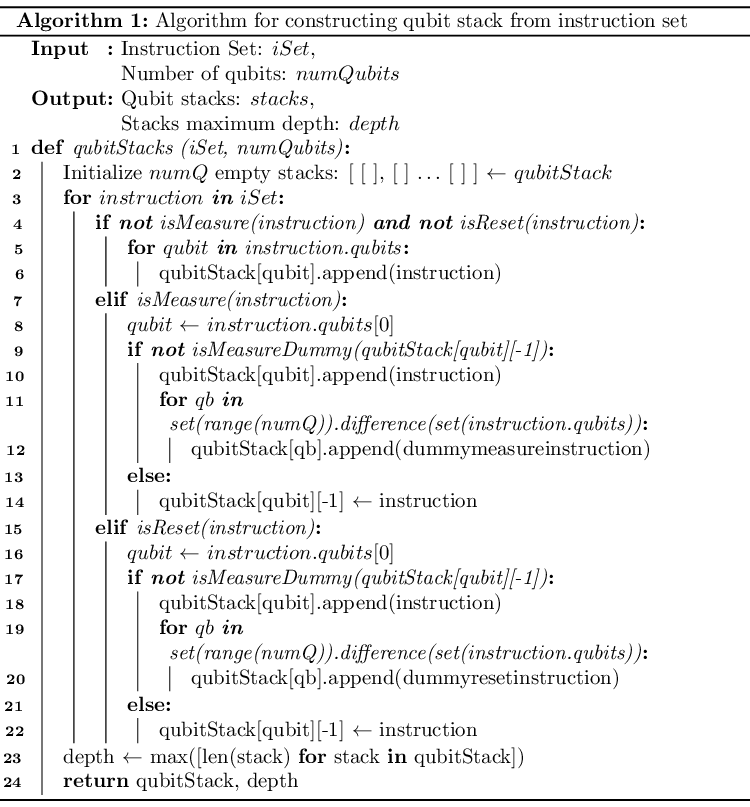}}
\caption{Pseudocode for the algorithm to construct the qubit operation
stack from the instruction set.}
\label{algo1}
\end{figure}

We implement a simple partitioning procedure, which may not be optimal,
but works well in practice.
Our first step is to construct a qubit operation stack from the instruction
set, which lists the temporal sequence of instructions that act on every qubit.
A multi-qubit instruction (e.g. C-NOT) is listed in the column of each
participating qubit.
In case of a single qubit measurement or reset, we add a dummy
instruction to the rest of the qubit columns as a barrier.
An example of this construction is shown in Fig.~\ref{circ_part}, and
the pseudocode of our algorithm is presented in Fig.~\ref{algo1}.

\begin{figure}
{\epsfxsize=7.9truecm \epsfbox{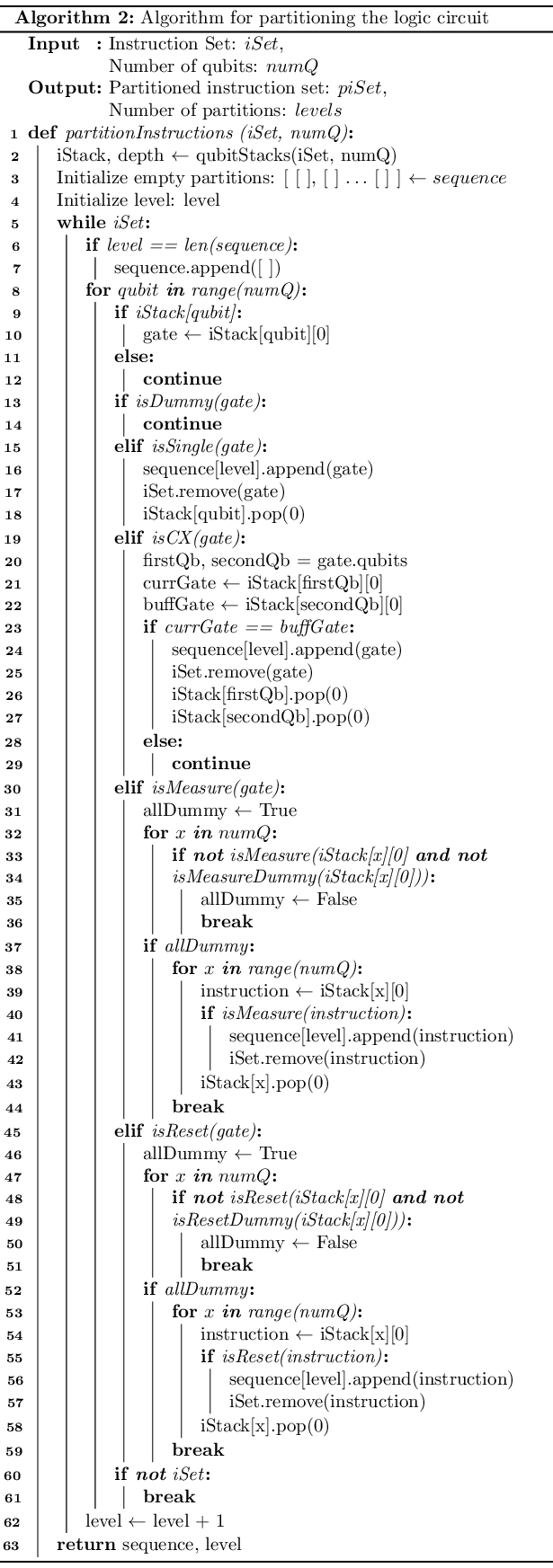}}
\caption{Pseudocode for the algorithm to partition the qubit operation
stack in to sequential levels.}
\label{algo2}
\end{figure}

Our next step is to sequentially inspect the bottom instruction for every
qubit column in the stack, and pop it in to a new partition under certain
conditions.
In case of a logic gate partition, at most only one instruction from a
column can go in to a partition, and a multi-qubit instruction can go in
to the partition only if it is present at the bottom of columns of each
participating qubit.
In case of a measurement partition, dummy instructions are ignored,
which allows simultaneous single qubit measurement or reset on distinct
qubits to be in the same partition (we call that partial measurement)
while preventing successive measurements on the same qubit from doing so.
This process of creating new partitions is repeated until all the columns
in the qubit operation stack become empty.
An example of this procedure is shown in Fig.~\ref{circ_part}, and
the pseudocode of our algorithm is presented in Fig.~\ref{algo2}.

At the end, we point out that an efficient software rescheduling of the
program instructions is desirable even when the algorithm is to be
implemented on a quantum hardware.

\begin{thebibliography}{99}
\bibitem{preskill}{J. Preskill,
        {Lecture Notes for the Course on Quantum Computation},
        \url{http://www.theory.caltech.edu/people/preskill/ph219/}}
\bibitem{nielsen}{M.A. Nielsen and I.L. Chuang, 
        {\it Quantum Computation and Quantum Information},
        (Cambridge University Press, 2000).}
\bibitem{NISQ}{J. Preskill, {Quantum Computing in the NISQ Era and Beyond},
        Quantum 2 (2018) 79.}
\bibitem{qcsimulators}{See for instance,
        \url{http://quantiki.org/wiki/list-qc-simulators}}
\bibitem{qiskit}{See \url{https://qiskit.org/} and\\
        \url{https://github.com/Qiskit/qiskit-terra}\\
        Qiskit has copyright under Apache License 2.0.}
\bibitem{coles}{See for instance, Abhijith J. et al.,
        {\it Quantum Algorithm Implementations for Beginners},
        {\tt arXiv:1804.03719v2} (2020).}
\bibitem{larose}{See for instance, R. LaRose,
        {\it Overview and Comparison of Gate Level Quantum Software Platforms},
        Quantum 3 (2019) 130.}
\bibitem{optcircuit}{See for instance,
        Y. Nam, N.J. Ross, Y. Su, A.M. Childs and D. Maslov, {\it Automated
        Optimization of Large Quantum Circuits with Continuous Parameters},
        npj Quantum Information 4 (2018) 23.}
\bibitem{optbasis}{H.-Y. Huang, R. Kueng and J. Preskill,
        {\it Information-theoretic bounds on quantum advantage in machine
        learning}, Phys. Rev. Lett. 126 (2021) 190505.}
\end{thebibliography}
\end{document}